\documentclass[fleqn,twoside]{article}
\usepackage{espcrc2}
\usepackage{amssymb}
\usepackage{bm}

\usepackage{graphicx}
\usepackage[figuresright]{rotating}
\newcommand{\AmS}{{\protect\the\textfont2
  A\kern-.1667em\lower.5ex\hbox{M}\kern-.125emS}}

\title{Electromagnetic neutrino-atom collisions: The role of electron binding}

\author{Konstantin A. Kouzakov\address{Department of Nuclear Physics and Quantum Theory of Collisions, Moscow State University, \\
        Leninskie Gory, Moscow 119991, Russia}\thanks{Email address: kouzakov@srd.sinp.msu.ru}
        and
        Alexander I. Studenikin\address{Department of Theoretical Physics, Moscow State University, \\
        Leninskie Gory, Moscow 119991, Russia}\address{Joint Institute for Nuclear Research, \\
        Dubna 141980, Moscow Region, Russia}\thanks{Email address: studenik@srd.sinp.msu.ru}
}

\begin{document}

\begin{abstract}
We present a new theoretical approach to neutrino-impact atomic
excitation and/or ionization due to neutrino magnetic moments. The
differential cross section of the process is given by a sum of the
longitudinal and transverse terms, which are induced by the
corresponding components of the force that the neutrino magnetic
moment imposes on electrons with respect to momentum transfer. In
this context, the recent theoretical studies devoted to the
magnetic neutrino scattering on atoms are critically examined.
\vspace{1pc}
\end{abstract}

\maketitle

\section{INTRODUCTION}
The Standard Model predicts the value of the neutrino magnetic
moment, in units of the Bohr magneton $\mu_B=e/(2m_e)$,
as~\cite{marciano77,lee77,fujikawa80}
\begin{equation}
\mu_\nu\approx3.2\times10^{-19}\frac{m_\nu}{1~{\rm eV}},
\end{equation}
where $m_e$ and $m_\nu$ are the electron and neutrino masses,
respectively. Any experimental evidence for a larger value of
$\mu_\nu$ will unequivocally indicate physics beyond the Standard
Model (a recent review of this subject can be found in
Ref.~\cite{giunti09}). The best upper limit for $\mu_\nu$ obtained
so far in experiments with reactor (anti)neutrinos is
$\mu_\nu\leq3.2\times10^{-11}$~\cite{gemma10} (see also references
in the review article~\cite{giunti09}). This is by an order of
magnitude larger than the most stringent astrophysical constraint
$\mu_\nu\leq3\times10^{-12}$~\cite{raffelt90}.

At small energy transfer $T$ the differential cross section (DCS)
for the magnetic neutrino scattering on a free electron (FE)
behaves as
$d\sigma_{(\mu)}/dT\propto1/T$~\cite{domogatskii71,vogel89}, while
that due to weak interaction, $d\sigma_{(w)}/dT$, is practically
constant in $T$~\cite{vogel89}. Therefore, one can enhance the
sensitivities of the reactor experiments by reducing the
low-energy threshold of the detectors in the deposited energy $T$.
However, the FE picture is applicable only when
$T\gg\varepsilon_b$, where $\varepsilon_b$ is the binding energy
of an atomic electron. If $T\sim\varepsilon_b$, the electron
binding effects must be taken into account. In this work, we
formulate a theoretical framework for the magnetic neutrino
scattering on atomic electrons which is similar to that developed
for the penetration of relativistic charged projectiles in
matter~\cite{fano63}. Within this approach the DCS is given by the
sum of two components due to the longitudinal and transverse
atomic excitations, respectively, which are induced by the
corresponding components of the force imposed by the neutrino
magnetic moment on electrons with respect to the direction of the
momentum transfer ${\bf q}$. It also enables us to clearly
distinguish between the contributions from excited atomic states
taken into account in Refs.~\cite{wong10} and~\cite{voloshin10},
respectively, and to inspect consistently the results of those
studies.

The article is organized as follows. Sec.~\ref{theory} delivers
general theory for the magnetic neutrino scattering on atomic
electrons as well as a critical account of the theoretical studies
published recently~\cite{wong10,voloshin10}. The conclusions are
drawn in Sec.~\ref{concl}. The units $\hbar=c=1$ are used
throughout unless otherwise stated.

\section{THEORY OF MAGNETIC NEUTRINO-IMPACT ATOMIC EXCITATION AND IONIZATION}
\label{theory}
We specify the incident neutrino energy and momentum by $E_\nu$
and ${\bf p}_\nu$, respectively. The atomic recoil is neglected
under the assumption $T\gg2E^2_\nu/M$, where $M$ is the nuclear
mass. The atomic target is supposed to be unpolarized and in its
ground state $|0\rangle$ with the corresponding energy $E_0$. We
treat the initial and final electronic systems nonrelativistically
under conditions $T\ll m_e$ and $\alpha Z\ll1$, where $Z$ is the
nuclear charge and $\alpha$ is the fine-structure constant. The
incident and final neutrino states are described by the Dirac
spinors assuming $m_\nu\approx0$. The neutrino electromagnetic
vertex associated with the neutrino magnetic moment is employed in
the low-energy limit
\begin{equation}
\label{el-m_vertex}\Lambda_{(\mu)}^i=\frac{\mu_\nu}{2m_e}\sigma^{ik}q_k,
\end{equation}
where $q$ is the virtual-photon four-momentum. Note that the
$\mu_\nu$ related contribution to the neutrino-atom scattering
couples neutrino states with different helicities and therefore it
does not interfere with that due to weak interaction.

Using first-order perturbation theory and the photon propagator in
the Coulomb gauge, the transition matrix element for the
considered process according to Eq.~(\ref{el-m_vertex}) is given
by~\cite{plb11}
\begin{eqnarray}
\label{matr_el}M_{fi}^{(\mu)}&=&\frac{2\pi\alpha\mu_\nu}{m_e|{\bf
q}|}(\bar{u}_{{\bf p}_\nu-{\bf q}\lambda_f}u_{{\bf
p}_\nu\lambda_i})\nonumber\\&{}&\times\Bigg\{\frac{2E_\nu-T}{|{\bf
q}|}\langle
n|\rho(-{\bf q})|0\rangle 
\nonumber\\
&{}&+\sqrt{\frac{(2E_\nu-T)^2-{\bf q}^2}{{\bf
q}^2-T^2}}\nonumber\\
&{}&\times\langle n|\hat{{\bf e}}_\perp\cdot{\bf j}(-{\bf
q})|0\rangle\Bigg\}, 
\end{eqnarray}
where $u_{{\bf p}\lambda}$ is the spinor amplitude of the neutrino
state with momentum ${\bf p}$ and helicity $\lambda$, $|n\rangle$
the final atomic state, and $\rho(-{\bf q})$ and ${\bf j}(-{\bf
q})$ the Fourier transforms of the electron density and current
density operators, respectively,
\begin{eqnarray}
\label{density}\rho(-{\bf q})&=&\sum_{j=1}^Ze^{i{\bf q}\cdot{\bf
r}_j}, \\ {\bf j}(-{\bf
q})&=&-\frac{i}{2m_e}\sum_{j=1}^Z\left(e^{i{\bf q}\cdot{\bf
r}_j}\nabla_j+\nabla_je^{i{\bf q}\cdot{\bf r}_j}\right),
\end{eqnarray}
and the unit vector $\hat{{\bf e}}_\perp$ is directed along the
${\bf p}_\nu$ component which is perpendicular to ${\bf q}$
($\hat{{\bf e}}_\perp\cdot{\bf q}=0$). Using Eq.~(\ref{matr_el}),
the DCS can be presented as 
\begin{equation}
\label{DCS}\frac{d\sigma_{(\mu)}}{dT}=\left(\frac{d\sigma_{(\mu)}}{dT}\right)_\parallel+\left(\frac{d\sigma_{(\mu)}}{dT}\right)_\perp,
\end{equation}
\begin{eqnarray}
\label{DCS_paral}\left(\frac{d\sigma_{(\mu)}}{dT}\right)_\parallel&=&\frac{\pi\alpha^2\mu_\nu^2}{m_e^2}
\frac{(2E_\nu-T)^2}{4E_\nu^2}\nonumber\\
&{}&\times\int_{T^2}^{(2E_\nu-T)^2}\left(1-\frac{T^2}{Q^2}\right)\nonumber\\&{}&\times
S(T,Q)\frac{dQ^2}{Q^2},
\end{eqnarray}
\begin{eqnarray}
\label{DCS_perp}\left(\frac{d\sigma_{(\mu)}}{dT}\right)_\perp&=&\frac{\pi\alpha^2\mu_\nu^2}{m_e^2}
\frac{(2E_\nu-T)^2}{4E_\nu^2}\nonumber\\ &{}&\times
\int_{T^2}^{(2E_\nu-T)^2}\left[1-\frac{Q^2}{(2E_\nu-T)^2}\right]\nonumber\\&{}&\times
R(T,Q)\frac{dQ^2}{Q^2},
\end{eqnarray}
where $Q=|{\bf q}|$ and
\begin{eqnarray}
\label{structure_factor}S(T,Q)&=&\sum_n|\langle n|\rho(-{\bf
q})|0\rangle|^2\nonumber\\&{}&\times\delta(T-E_n+E_0),
\\\label{structure_factor_cur} R(T,Q)&=&\sum_n|\langle n|\hat{{\bf
e}}_\perp\cdot{\bf j}(-{\bf
q})|0\rangle|^2\nonumber\\&{}&\times\delta(T-E_n+E_0).
\end{eqnarray}
The sums in Eq.~(\ref{structure_factor}) run over all atomic
states $|n\rangle$, with $E_n$ being their energies, and, since
the ground state $|0\rangle$ is unpolarized, do not depend on the
direction of ${\bf q}$.

The longitudinal term~(\ref{DCS_paral}) is associated with atomic
excitations induced by the force that the neutrino magnetic moment
exerts on electrons in the direction parallel to ${\bf q}$. The
transverse term~(\ref{DCS_perp}) corresponds to the exchange of a
virtual photon which is polarized as a real one, that is,
perpendicular to ${\bf q}$. It resembles a photoabsorption process
when $Q\to T$ and the virtual-photon four-momentum thus approaches
a physical value, $q^2\to0$. Due to selections rules, the
longitudinal and transverse excitations do not interfere (see
Ref.~\cite{fano63} for detail).

The properties of Eq.~(\ref{DCS_paral}) were studied in the
work~\cite{voloshin10}, where the transverse component was
unaccounted. It is determined by the dynamical structure factor
\begin{equation}
S(T,Q)=\frac{1}{\pi}{\rm Im}F(T+E_0,Q),
\end{equation}
where the density-density Green's function $F$ is
\begin{eqnarray}
\label{green}
F(E,Q)&=&\langle0|\rho({\bf q})\frac{1}{E-H-i0}\rho(-{\bf
q})|0\rangle\nonumber\\
&=&\sum_n\frac{|\langle n|\rho(-{\bf q})|0\rangle|^2}{E-E_n-i0},
\end{eqnarray}
with $H$ being the atomic Hamiltonian. In Ref.~\cite{voloshin10}
the dispersion relation for the function $F$ was formulated as
\begin{equation}
\label{disp_rel_voloshin}F(E,Q)=\frac{1}{\pi}\int_0^\infty\frac{{\rm
Im}F(E,Q')}{Q'^2-Q^2-i0}dQ'^2.
\end{equation}
Consider the limit $Q\to0$. The electron density
operator~(\ref{density}) at ${\bf q}=0$ is by definition
$\rho(0)=Z$ and hence
\begin{eqnarray}
F(E,0)&=&\langle0|\rho(0)\frac{1}{E-H-i0}\rho(0)|0\rangle\nonumber\\
&=&\frac{Z^2}{E-E_0}.
\end{eqnarray}
Using it in Eq.~(\ref{disp_rel_voloshin}) when $Q=0$, we arrive at
the sum rule~\cite{plb11}
\begin{equation}
\label{voloshin_sum_rule}\int_0^\infty
S(T,Q)\frac{dQ^2}{Q^2}=\frac{Z^2}{T}.
\end{equation}
Note that the value of $F(T+E_0,0)$ is calculated in
Ref.~\cite{voloshin10} erroneously, namely as
\begin{equation}
F(E,0)=\frac{Z}{E-E_0}.
\end{equation}
Following the procedure of Ref.~\cite{voloshin10}, which implies
the use of Eq.~(\ref{voloshin_sum_rule}) for evaluating the
integral in Eq.~(\ref{DCS_paral}) under assumptions of small $T$
and large $E_\nu$, we arrive at the result where the factor of
$Z^2$ occurs. This means that the atomic effects result in a
coherent enhancement of the DCS as compared to the case of $Z$
free electrons, where a typical incoherent-scattering factor of
$Z$ is encountered (the same as, for instance, in the Compton
scattering). This conclusion is not consistent with the incoherent
character of the considered inelastic scattering process.

Taking into account that ${\bf j}=(\rho{\bf v}+{\bf v}\rho)/2$,
with ${\bf v}$ being the electron velocity operator, we can
estimate the ratio of the functions~(\ref{structure_factor_cur})
and~(\ref{structure_factor}) as $\sim\upsilon^2_a$, where
$\upsilon_a\ll1$ is a characteristic velocity of atomic electrons.
Therefore, one might expect the transverse component to play a
minor role in Eq.~(\ref{DCS}). However, the authors of
Ref.~\cite{wong10} came to the contrary conclusion that this
component strongly enhances due to atomic ionization when
$T\sim\varepsilon_b$. The enhancement mechanism proposed in
Ref.~\cite{wong10} is based on an analogy with the photoioniztion
process. As mentioned above, when $Q\to T$ the virtual-photon
momentum approaches the physical regime $q^2=0$. In this case, we
have
%
\begin{equation}
\label{photoeffect}\left.\frac{R(T,Q)}{Q^2}\right|_{Q\to
T}=\frac{\sigma_\gamma(T)}{4\pi^2\alpha T},
\end{equation}
where $\sigma_\gamma(T)$ is the photoionization cross section at
the photon energy $T$~\cite{akhiezer_book}. The limiting
form~(\ref{photoeffect}) was used in Ref.~\cite{wong10} in the
whole integration interval. Such a procedure is incorrect, for the
integrand rapidly falls down as $Q^2$ ranges from $T^2$ up to
almost $4E_\nu^2$, especially when $Q\gtrsim r_a^{-1}$, where
$r_a$ is a characteristic atomic size (within the Thomas-Fermi
model $r_a^{-1}=Z^{1/3}\alpha m_e$~\cite{landafshiz_book}). This
fact reflects a strong departure from the real-photon regime.
Thus, we can classify the enhancement of the DCS claimed in
Ref.~\cite{wong10} as spurious. It should be noted in this
connection that when the present work (as well as~\cite{plb11})
had already been completed and submitted for publication the
authors of Ref.~\cite{wong10} had disproved their claim (see
Ref.~\cite{wong10_arxiv} for detail).

\section{CONCLUSIONS}
\label{concl}
To summarize, we have performed a theoretical analysis of the
magnetic neutrino scattering on atomic electrons. For this purpose
we have divided the DCS into two components corresponding to the
longitudinal and transverse atomic excitations. This allowed us to
demonstrate in a physically transparent fashion the deficiencies
of the recent theoretical predictions concerning the role of the
atomic effects in the magnetic neutrino scattering~\cite{wong10}.
No enhancement mechanism due to electron binding effects has been
determined~\cite{plb11} (see also Ref.~\cite{voloshin10}), in
contrast to Ref.~\cite{wong10}. At the same time, the attempt to
argue the insignificance of the atomic effects by means of
analytical calculations~\cite{voloshin10} needs further
elaboration~\cite{plb11,ksv}.

Finally, it is unreasonable to expect the effects of atomic
excitation and/or ionization to introduce enhancement of the
sensitivities of the experiments searching for neutrino magnetic
moments. In this respect, it will be interesting to explore the
role of coherent magnetic neutrino scattering on atoms in
detectors, which case, however, requires much lower energy
thresholds in the deposited energy $T$ ($\sim100$~eV) than
presently attainable in the detectors ($\sim1$~keV).

\section*{ACKNOWLEDGEMENTS}
We are grateful to Mikhail B. Voloshin for useful discussions and
valuable comments. One of the authors (A.I.S.) is thankful to
Gianluigi Fogli and Eligio Lisi for invitation to attend NOW2010.
The work of K.A.K. (in part) and A.I.S. is supported by RFBR grant
11-02-01509-a. K.A.K. also acknowledges partial support from RFBR
grant 11-01-00523-a.

\end{document}